\documentclass[%
 aip,
 amsmath,amssymb,
 reprint,%
]{revtex4-1}

\usepackage{graphicx}
\usepackage{dcolumn}
\usepackage{bm}

\usepackage[utf8]{inputenc}
\usepackage[T1]{fontenc}
\usepackage{mathptmx}
\usepackage{gensymb}
\usepackage{tablefootnote}
\usepackage{caption}
\usepackage{subcaption}
\usepackage{nomencl}
\usepackage{amssymb}
\usepackage{parskip}
\usepackage{setspace}
\usepackage{titlesec}
\makenomenclature
\usepackage{etoolbox}
\usepackage{adjustbox}
\renewcommand\nomgroup[1]{%
  \item[
  \ifstrequal{#1}{A}{Symbols}{%
  \ifstrequal{#1}{B}{Greek symbols}{%
  \ifstrequal{#1}{L}{Superscripts}{%
  \ifstrequal{#1}{U}{Subscripts}{}}}}%
]}

\usepackage[marginal]{footmisc}

\begin{document}
\nocite{*} 
\preprint{AIP/123-QED}
\title[A Probability Space at Inception of Stochastic Process]{A Probability Space at Inception of Stochastic Process}

\author{Liteng Yang$^1$}
\author{Yuliang Liu$^1$}
\author{Jing Liu$^2$}
\author{Hongxuan Li$^2$}
\author{Wei Chen$^*$}%

\thanks{1,2 contribute equally, *corresponding author: chenwei23@ustb.edu.cn}

\begin{abstract}
Recently, progress has been made in the theory of turbulence, which provides a framework on how a deterministic process changes to a stochastic one owing to the change in thermodynamic states. It is well known that, in the framework of Newtonian mechanics, motions are dissipative; however, when subjected to periodic motion, a system can produce nondissipative motions intermittently and subject to resonance. It is in resonance that turbulence occurs in fluid flow, solid vibration, thermal transport, etc. In this, the findings from these physical systems are analyzed in the framework of statistics with their own probability space to establish their compliance to the stochastic process. In particular, a systematic alignment of the inception of the stochastic process with the signed measure theory, signed probability space, and stochastic process was investigated. It was found that the oscillatory load from the dissipative state excited the system and resulted in a quasi-periodic probability density function with the negative probability regimes. In addition, the vectorial nature of the random velocity splits the probability density function along both the positive and negative axes with slight asymmetricity. By assuming that a deterministic process has a probability of $1$, we can express the inception of a stochastic process, and the subsequent benefit is that a dynamic fractal falls on the probability density function. Moreover, we leave some questions of inconsistency between the physical system and the measurement theory for future investigation. We believe that the establishment of the probability density function of resonance nondissipative dynamics in contemporary statistics should make many mathematical tools available and the analytical formulas for the random velocity and probability density function can provide a convenient platform for the development of statistics.
\end{abstract}

\maketitle

\section{\label{sec:level1}Introduction}
Around us, many phenomena happen randomly and the continuous evolution of those random phenomena are known as the stochastic processes \cite{1,2,3}. In the framework of modern physics and mathematics, any stochastic process can only be induced by a stochastic variable or another stochastic process. Two plausible routes from deterministic to stochastic processes are instability and bifurcation, both of which suffer operability issues in practical scenarios. Mathematically, instability conditions in time and space are rarely encountered, and their effects are very limited \cite{4,5}; however, the stochastic process usually fills the entire space of the domains. The discovery of the first and second Feigenbaumer constants implies that every bifurcation takes time \cite{6}; hence, it consumes a finite amount of time in the transition from deterministic to stochastic processes for all bifurcating steps, contradicting observations where the inception of a stochastic process is nearly instant \cite{7}. Recently, a physical process has been identified in which a deterministic, dissipative process with periodic motion generates a stochastic, nondissipative process \cite{8}. When it is used to describe any deterministic, dissipative process, there are topologies of nullified entropy in time and space with periodic motions, which follow the principles of nondissipative dynamics, which is completely different from the laws of dissipative motion. Moreover, the excitation of nondissipative dynamics has a stochastic component. In this article, we briefly describe the physical process of the transition of thermodynamic states and the nondissipative stochastic process, and then focus on the properties of its probability distribution function. There is sufficient evidence that this stochastic process is ubiquitous, and therefore, it is highly desirable to further understand this new nondissipative process, stochastic process. We believe that insights into the new stochastic process will help in research, application, augmentation and detection in a wide range of scientific and engineering settings.

Randomness refers to the nature of life. For example, when you take a bus to go to work or school, you would not know exactly when it arrives at school and how long it takes since the events along the way randomly interrupt the school bus. The example elucidates two known facts: first, a random process is promoted by random processes within; second, the magnitude of deviation in a random process depends on the magnitude of each participating random process and the relation between the main process and its subprocesses. The first observation is the foundation of modern probabilistic theory and stochastic processes, where any randomness is captured by the introduction of a measurable sample space $\left(\Omega,\mathcal{F}\right)$, where probability measures can be imposed. Hence, a stochastic process is a collection of random variables $X=\left\{X_t;0\leq t <\infty\right\}$ on $\left(\Omega,\mathcal{F}\right)$, which takes values in a second measurable space $\left(\mathcal{S},\mathcal{T}\right)$, named the state space, and is a non-trivial collection. The state space $\left(\mathcal{S},\mathcal{T}\right)$ is defined as the d-dimensional Euclidean space with the Borel sets and $\sigma$-field, denoted as $S=\mathbb{R}^d$, $\mathcal{T}=\mathfrak{B}(U)$, which is the smallest $\sigma$-field containing all open Borel sets of a topological space $U$. In measurement theory, the topological sets of the state space are always nontrivial. Therefore, by definition, any stochastic process contains at least one random variable $X=\left\{X_t;0\leq t <\infty\right\}$. Thus far, there is no known stochastic process that does not rely on random variables to trigger the processes of randomness. In other words, a deterministic process does not generate a stochastic process without a borehole set and $\sigma$-fields \cite{1,9}. To accommodate a wide range of observations on an arbitrary deterministic process transitioning to a stochastic process, a Brownian process or Wiener process is mathematically adopted by Ito's calculus, where the consistent derivative and integral forms of a stochastic process to those of the Leibniz calculus are essential \cite{2}. 

It is challenging to allow deterministic governing differential equations to generate stochastic responses through the introduction of Brownian motions and Ito calculus, although it heads to that direction \cite{10,11}. The mathematics of measurement theory and stochastic differential equations may have laid down the foundation and were preparing for that \cite{3,12}. In fact, with or without the concern of the inception of a stochastic process, the responses of the deterministic governing differential equations to Brownian motion or a generic stochastic process should be investigated. The treatment of randomness of field variables and background noises has been separated from the averaged or mean field by an independent stochastic differential equation, and the benefit of doing so is that if bulk fields are in a steady state, they can be calculated fairly straightforwardly with the established method. There are different versions of how the mean and noise are included and formulated, and the results produced by them bear many similarities, although they deviate somewhat. For example, the burst of a random quantity and the start of a stochastic process cannot be described or explained using the approach above. It is more likely that the onset of the random process shares the same physical ground as its transport, that is, the isentropic process for acoustic noises, which coincidentally aligns with their sustainability and persistence; otherwise, random motions are consumed and dissipated quickly. Additionally, we find a mathematical theory and governing differential equations that elucidate the life cycle of a stochastic process and the interactions of deterministic and stochastic processes. 

Recent progression turbulence has shown that random fluctuations in thermal propagation, fluid flow, and solid elastic deformation can be explained by nondissipative dynamics \cite{8}, where the local, instantaneous thermodynamic state changes from dissipative to nondissipative. Subsequently, motions must follow the nondissipative governing differential equations instead of the dissipative equations, where the local entropy mass density s is introduced in energy conservation: \cite{13},
\begin{equation}
\rho T\frac{Ds}{Dt} =-\mathbf{\nabla}\cdot\mathbf{q}+\mathbf{\tau}:\nabla\mathbf{u}
\label{1eq}
\end{equation}
where $\mathbf{\tau}$, $\mathbf{q}$, $\mathbf{u}$, $\rho$ and T are the stress tensor, heat-flux vector, velocity vector, density and temperature, respectively. According to the second law of thermodynamics, when s=0, the thermodynamic state is isentropic or non-dissipative and we obtain a local non-dissipative material parcel, which can be either fluid or solid. Under isentropic conditions, the equations of motion follow the Euler equation:
\begin{align}
&\frac{D\rho}{Dt}+\rho\mathbf{\nabla}\cdot{\mathbf{u}} = 0\label{2eq}\\
&\rho\left(\frac{{\partial \mathbf{u}}}{{\partial t}}+\mathbf{u}\cdot\nabla{\mathbf{u}}\right) = -\nabla{p} \label{3eq}\\
 &\rho \frac{De}{Dt} =-\left(\mathbf{\nabla}\cdot pv\right) \label{4eq}
\end{align}
Utilizing of (\ref{2eq}) and (\ref{3eq}), we can derive the wave equation for pressure and take the derivative of (\ref{4eq}) to time and space respectively, to obtain the wave equation for temperature:
\begin{equation}
     \frac{\partial^2p}{\partial t^2}-\left(\frac{\partial{p}}{\partial{\rho}}\right)_s\nabla^2{p} =\Psi \label{5eq}
\end{equation} 
\begin{equation}
   \frac{\partial^2 T}{\partial t^2}-\mathbf{u}\cdot\mathbf{u}\nabla^2T =\Phi  \label{6eq}
\end{equation}
where $\Psi$ and $\Phi$ are functions of velocity and temperature, respectively. Treating (\ref{5eq}) and (\ref{6eq}) as a two-parameter dynamical system and employing a small perturbation analysis, we obtain the nondissipative resonant governing differential equations:
\begin{equation}\label{7eq}
\left\{ \begin{array}{ll}  
\begin{aligned}
\rho\left(\frac{\partial\mathbf{u}}{\partial{t}}\pm\mathbf{c_1}\cdot\nabla{\mathbf{u}}\right) &= -\nabla{p}\\
\rho c_v\left(\frac{\partial{T}}{\partial{t}}\pm\mathbf{c_2}\cdot\nabla{T}\right) &= 0
\end{aligned}
 \end{array}\right.
\end{equation}
where $\mathbf{c_1}=c_1 \mathbf{i}+c_1 \mathbf{j}+c_1 \mathbf{k}$ and $\mathbf{c_2}=c_2 \mathbf{i}+c_2 \mathbf{j}+c_2 \mathbf{k}$, $c_1$ and $c_2$ are the speeds of the first and second sounds, respectively. The solutions of (\ref{7eq}) are the velocity and temperature of the material parcel at resonant, nondissipative states, which are excited with broad band frequencies in time and space; hence, the solutions possess the characteristic traits of stochastic processes and the field variables attain and appear to be random. In other words, from (\ref{1eq})-(\ref{7eq}), we show the physical steps of how a deterministic process becomes a stochastic process and the inception of a stochastic process.

The mathematical derivations from (\ref{1eq}) to (\ref{7eq}) briefly summarizes the logics and reasoning behind them, which are adopted only with sufficient observations and validations from practical and outstanding problems. In what follows, the mathematical embodiment of nondissipative dynamics addresses critical questions that may arise from fundamental physics and observations. The application of entropy or excess entropy and the difference in the local entropy mass density of the dissipative and nondissipative entropy may be considered as the nonequilibrium thermodynamics of the second law of the local form, which has long been sought for \cite{14,15}. The fundamental meaning of local entropy is the partitioning of the domain into two mutually exclusive domains, with and without energy loss. Domains without energy loss have also been called collective motion, superfluid, and nondissipative motion. To visualize such a partition of thermodynamic states, the concept of Hausdorff space ($H$) can be introduced; the nondissipative parcel is also a Hausdorff space $(h\in H)$, where the nondissipative state shares the disjoint space $(a_1\in A_1 )$, also called extensions, and the dissipative state occupies the joint space $(a_2\in A_2)$, also called absolutes, where $a_1\cup a_2=0$ and $a_1\cap a_2=h$. Depending on the excitation state, the nondissipative state may take the joint space instead of the disjoint space. When the nondissipative state surrounds a solid parcel, although there are materials in the dissipative state in the solid parcel, the entire parcel can still behave in an entirely nondissipative state and move nondissipatively. This topological structure is maintained in the domain that generates the small parcels, and the domain contains the small parcels recursively.

On the inception of a stochastic process from a deterministic process, the rebuttal may be expressed that the perturbation of the dissipative state should be treated as the initial stochastic process instead of a deterministic one. In view of the theory of excess entropy, the initial perturbation can be considered either a stochastic process or simply a regular wave because any arbitrary perturbation can be decomposed into many sinusoidal waves in the sense of Fourier transformation. When one wavelet participates in a motion on the dissipative side, the local entropy approaches zero spatiotemporally and kicks off the non-dissipative state and the corresponding solutions of velocity and temperature are readily obtained in the non-dissipative state. Thereafter, the probability function of the field variables is derived, which does not rely on the functionality and manner of the initial functions. The explanation also clarifies another similar but less obvious rebuttal, which claims that the employment of local entropy or local excess entropy implicitly introduced a stochastic process because of the strong affiliation of entropy to chaos and randomness. The local entropy, by definition, should be considered as a measure of energy conversion into heat from any other kind of energy because thermal energy is viewed as the random motion of metaphase, which means that a new stochastic process has been created through the perturbations. If the thermodynamic state is not surveyed throughout each period, the excitation and random motion may be mistakenly considered to generated in the perturbation. Now, as described through (\ref{1eq}) to (\ref{7eq}), this randomness is created at a fraction of time in a period. In other words, the nondissipative motion fits that description as a consequence. On the other hand, if the wavelets and excitation induce it, the oscillation in the dissipative state obeys Fourier’s law, dissipates quickly, and then stops the oscillation. Because the observation tells us otherwise, it only makes sense that the stochastic process is thermodynamically in the nondissipative state.

An occasional brush in the theory of nondissipative dynamics is the appearance of ensembles of classical mechanics. The dash may rest on the fact that some of the differential equations and technical terms, such as nondissipative motion, have appeared elsewhere. It is understandable that similar terms here may cause confusion, which should be avoided if proper definitions and explanations are provided. Moreover, development is most likely to change, adapt, and deform the framework, contents, and embodiments of the theory, which makes little value to devote effort to the terms. However, since classical mechanics have been validated with countless practical cases, it should leave us with tremendous confidence and enlightenment when a new physics has been exposed after further bisection and reassembling them in a different way. In a way, it should keep all of us humble to what we have inspired from past intellect.

In this study, we first briefly demonstrate the derivation of a probability function and its parameters. Second, we discuss the properties of the probability density function. Third, examples of these applications are provided. Finally, we summarize our findings.

\section{\label{sec:level2}Probability Function}
The objective of this section is to describe the essential steps in deriving the probability density function. The process starts from the general solution of the local velocity in resonance for the fluid and the local strain rate of deformation for solids. On the dissipative side, owing to the transition of thermodynamic states, the impulse causes oscillation in excitation on the nondissipative material, which results in the resonant velocity as a stochastic variable, u, on time and space. The corresponding initial and boundary conditions on the boundary of the second law are defined by the conservation of the mass, momentum and energy. The probability function was calculated from the conditional correlation between the resonant velocity and temperature. The probability distribution can be viewed in both time and space. The invariance of the temporal part of the distribution was first discovered by Taylor in $1938$, which denotes turbulence fluctuations \cite{16}. Later, it was found that the correlation function or probability density function can be considered invariant regardless of the nature of the random variables.

From (\ref{7eq}), if we only consider the velocity part, the equation of motion can be written as
\begin{equation}
\rho\left(\frac{\partial u_i}{\partial{t}}+c_1\frac{\partial u_j}{\partial{x_j}}\right) = -\frac{\partial p}{\partial{x_i}}\label{8eq},i=1,2,3
\end{equation}
where $j$ follows tensor contraction for summation. Apply the separation method to the velocity components $u_i$,
\begin{equation}
u_i=T_i(t)X_i(x)Y_i(y)Z_i(z)\label{9eq}
\end{equation}
With (\ref{9eq}), (\ref{8eq}) becomes
\begin{equation}
\frac{1}{c_1}\frac{1}{T_i}\frac{dT_i(t)}{dt}+\frac{1}{X_i}\frac{d   X_i(x)}{dx}+\frac{1}{Y_i}\frac{Y_i(y)}{dy}+\frac{1}{Z_i}\frac{dZ_i(z)}{dz} 
= -\frac{1}{\rho u_i}\frac{\partial p}{\partial x_i} \label{10eq}
\end{equation}
The eigenfunction of (\ref{10eq}) is given
\begin{equation}
n_1+n_2+n_3+n_4=n_0=-\frac{1}{\rho u_i}\frac{\partial p}{\partial x_i}
\label{11eq}
\end{equation}
where $n_l=(\pm\alpha_l+if_l)k_l,l=0,1,2,3,4$ are complex constants. Because it is only a function of time, the first term on the left of (\ref{10eq}) must equate to a constant $n_1=\frac{1}{c_1}\frac{1}{T_i}\frac{dT_i(t)}{dt}$ if (\ref{8eq}) and (\ref{9eq}) hold true. Similarly, $n_2$, $n_3$ and $n_4$ are equal to the corresponding terms in Eq.(\ref{10eq}) for all three coordinates. From the left side of equation (\ref{11eq}), the summation is naturally equal to another constant $n_0$ in order to maintain equation(\ref{10eq}), which results in the equation
\begin{equation}
n_0=-\frac{1}{\rho u_i}\frac{\partial p}{\partial x_i}\label{12eq}
\end{equation}
Solve (\ref{12eq}) by integration, $p=-n_0T_i(t)Y_i(y)Z_i(z)\int X_i(x)dx$. From (\ref{9eq}), (\ref{11eq}) and (\ref{12eq}), the general solution in the direction of $Z$ Z-coordinate can be written as
\begin{equation}
\begin{aligned}
    u_z =u_0 &\left[\sum\nolimits_{k_1}^{K_1}\left(A_{1,k_1}cosc_1f_{k_1}k_1t  +A_{2,k_1}sinc_1f_{k_1}k_1t\right)e^{-c_1\alpha_{k_1}k_1t}\right]\\ &\left[\sum\nolimits_{k_2}^{K_2}\left(B_{1,k_2}cosf_{k_2}k_2x+B_{2,k_2}sinf_{k_2}k_2x \right)e^{-\alpha_{k_2}k_2x}\right]\\    &\left[\sum\nolimits_{k_3}^{K_3}\left(C_{1,k_3}cosf_{k_3}k_3y+C_{2,k_3}sinf_{k_3}k_3y \right)e^{-\alpha_{k_3}k_3y}\right]\\      &\left[\sum\nolimits_{k_4}^{K_4}\left(D_{1,k_4}cosf_{k_4}k_4z+D_{2,k_4}sinf_{k_4}k_4z \right)e^{-\alpha_{k_4}k_4z}\right]\\ 
    \label{13eq}
\end{aligned}
\end{equation}
We choose the velocity in the z direction in (\ref{13eq}).

To solve the integral constants in (\ref{13eq}), we must employ the concept developed by the so-called boundary of the second law \cite{8}, which determines the instantaneous and spatial locations where the entropy is zero but the velocity has a limited value. In other words, the boundary of the second law is defined as that in dynamics, where the local entropy or excess entropy becomes zero instantaneously; hence, according to the second law of thermodynamics, the instants and locations with zero entropy have to follow the governing differential equations of nondissipative motion. To maintain conservation, the velocity from the dissipative side must be equal to that of the nondissipative side \cite{17},
\begin{equation}
u_z^{\dagger^2}(t,x,y,z)=u_z\label{14eq}
\end{equation}
where $u_z^{\dagger}$ is the velocity at time $t$ and location $(x,y,z)$. From the boundary and initial conditions established from (\ref{14eq}), we obtain the integral constants,
\begin{equation}\label{eq:15}
\left\{
\begin{aligned}
A_{1,k_1} &= \frac{\alpha_{k_1}}{f_{k_1}}, & A_{2,k_1} &= 2, \\
B_{1,k_2} &= \frac{\alpha_{k_2}}{f_{k_2}}, & B_{2,k_2} &= 2, \\
C_{1,k_3} &= \frac{\alpha_{k_3}}{f_{k_3}}, & C_{2,k_3} &= 2, \\
D_{1,k_4} &= \frac{\alpha_{k_4}}{f_{k_4}}, & D_{2,k_4} &= 2.
\end{aligned}
\right.
\end{equation}
Owing to the sudden change in the governing differential equations, the velocity on the boundary of the second law (\ref{14eq}) is best represented by the delta functions, which describe the physical process in view of the nondissipative side. Before the change in the thermodynamic state, there is no physical existence of nondissipative material in the given domain. At the time of change, a parcel of nondissipative material appears to have a velocity of $u_z^{\dagger}(t,x,y,z)$, which continues to excite the parcel of the nondissipative material, which follows the partial differential equations, for example for the velocity of the temporal component \cite{17},
\begin{equation}
u_z^{\dagger}(t)=\sum\limits_{l}^{L}\sum\limits_{k_1}^{K_1}u_{z_0}\delta\left[2\pi c_1k_1f_1t-l\pi\right]\label{16eq}
\end{equation}
where $l$ represents a train of the delta function, also called a Dirac comb, Dirac train or Shah comb. Velocity (\ref{eq:15}) includes any combination of $10$ elementary profiles for the fluid and several more elementary profiles for the solid, which leaves the velocity field (\ref{13eq}) as random fields.

The physical process in a thermodynamic transition from a dissipative one to a non-dissipative one marks a deterministic process becoming a stochastic process. Before the transition, the motion is described by governing partial differential equations with the corresponding constitutive relations, which have been considered deterministic in mathematics. The superposition of disturbances from small waves is also deterministic, because they both obey the same governing partial differential equations. Because nondissipative motions follow the governing differential equations of resonant and nondissipative dynamics, the velocity field (\ref{13eq}) will certainly be different from that of dissipative motion. In a dissipative velocity field, dissipation and friction impede the motion of sinusoidal perturbations, and fluctuations die down quickly from viscous forces.In contrary, the motion in nondissipative dynamics does not decay but continues to follow the solution (\ref{eq:15}), which sustains the motion for a very long time. Hence, the superposition of all velocity components becomes stochastic and random. Subsequently, we demonstrate that the velocity field possesses many properties of a stochastic process.

 A common postulate on nonlinearity has been circulated on the inception of a stochastic process, which is based on the observation that when a process gradually approaches equilibrium or near-equilibrium, an apparently new process starts with a stochastic process. The observation has been categorized as nonlinear because this new process and the emergence of the stochastic process can occur at the minimum disturbances. The rebuttal on the validity of Equation (\ref{7eq}) is on the inconsistency of the nonlinearity of the stochastic process vs. the linearity of Equation (\ref{7eq}). Here, we define and discern the nonlinearity of the phenomenon. There are two types of nonlinearity. The first type of nonlinearity is global nonlinearity, which shows the nonlinearity when the domains become large, which is not the nonlinearity that we are interested in here. In the second type, nonlinearity begins to appear when the domains become smaller. Here, we briefly prove the nonlinearity of the phenomenon and show that the nonlinearity of a phenomenon has nothing to do with the linearity of the partial differential equations.

 Conventionally, two methods have been used to prove the nonlinearity of a phenomenon. The Taylor series can be used in a functional analysis of the phenomenon. Second, if it preserves the linearity by adding or subtracting one process from the other, the superposition of multiple linear processes is still linear; otherwise, it is nonlinear. As there is only a qualitative description, we will develop a more rigorous method to define a nonlinear phenomenon. First, a phenomenon $\frak{P}$ is defined as a collection of various sets and objects in the Sobolev Space and complex variables:
\begin{equation}
\frak{P}=\Sigma_{\beta}\label{17eq}
\end{equation}
where $\sum$ represents all physical laws, or represents partial and ordinary differential equations and $\beta$ are the corresponding sets of initial and boundary conditions, constraints, physical limits, and so on. Adopt the symbolic expression in a small variation of the constraints and boundary conditions by $\Delta\beta=\beta_2-\beta_1$ and the corresponding changes in the solutions along all physical laws by $\Delta\Sigma=\Sigma_{\beta_2}-\Sigma_{\beta_1}$, which leads to a corresponding change in the phenomenon. 
\begin{equation}
\Delta\frak{P}=\Delta\Sigma_{\Delta\beta}=\Sigma_{\beta_2}-\Sigma_{\beta_1}\label{18eq}
\end{equation}
If the phenomenon of interest is completely dissipative, $\mathfrak{D}$, the two types of motion are represented by two infinitesimal waves $W_1$ and $W_2$. Under condition $\beta_1$, phenomenon (18) can be written as,
\begin{equation}
\Delta\mathfrak{D}=\Sigma_{\beta_1}\left(W_1+W_2\right)=\Sigma_{\beta_1}\left(W_1\right)+\Sigma_{\beta_1}\left(W_2\right)\label{19eq}
\end{equation}
where we can linearize the physical laws $\Sigma_\beta$ to satisfy (\ref{19eq}), that is, the superposition.From (\ref{18eq}), superposition (\ref{19eq}) can be further written as
\begin{equation}\label{20eq}
\begin{aligned}
\Delta\mathfrak{D}&=\Delta\Sigma_{\Delta\beta}=\Sigma_{\beta_2}-\Sigma_{\beta_1}\\
&=\Sigma_{\beta_2}\left(W_1\right)+\Sigma_{\beta_2}\left(W_2\right)-\Sigma_{\beta_1}\left(W_1\right)+\Sigma_{\beta_1}\left(W_2\right)\\
&=\Delta\Sigma_{\Delta\beta}\left(W_1\right)+\Delta\Sigma_{\Delta\beta}\left(W_2\right)
\end{aligned}
\end{equation}
This phenomenon follows the superposition principle. Similarly, we can prove that the completely nondissipative phenomenon $\mathfrak{N}$ satisfies (\ref{20eq}).
\begin{equation}
\Delta\mathfrak{N}=\Delta\Sigma_{\Delta\beta}\left(X_1\right)+\Delta\Sigma_{\Delta\beta}\left(X_2\right)\label{21eq}
\end{equation}
where $X_1$ and $X_2$ are two excitations due to the impulsive passage of velocity from dissipative to nondissipative motions. The phenomena in (\ref{20eq}) and (\ref{21eq}) have the following limitations:
\begin{equation}
\lim_{\Delta\beta \to 0}\Delta\mathfrak{D}=0 \text{ and }\lim_{\Delta\beta \to 0}\Delta\mathfrak{N}=0\label{22eq}
\end{equation}
where (\ref{22eq}) means that the dissipative and nondissipative dynamics follow the superposition such that they are linear independently. If the two phenomena are added without the transition from dissipative to nondissipative states, we have
\begin{equation}
\mathfrak{B}=\mathfrak{D}+\mathfrak{N} \text{ and }\Delta\mathfrak{B}=\Delta\mathfrak{D}+\Delta\mathfrak{N}\label{23eq}
\end{equation}
From (\ref{20eq}), (\ref{21eq}), and (\ref{22eq}), the phenomenon is linear owing to the superposition of (\ref{20eq}) and (\ref{21eq}). However, if we include the transition from dissipative to nondissipative processes, the functional analysis yields,
\begin{equation}
\mathfrak{B}=\mathfrak{D}+\mathfrak{N}+\Sigma_i\delta\left[t_i,\mathbf{X_i}\right]\label{24eq}
\end{equation}
where $\Sigma_i\delta\left[t_i,\mathbf{X_i}\right]$ is the sum of all delta function at each instant and location $i$ due to the impulses on the boundary of the second law. The superposition of (24) yields,
\begin{equation}
\Delta\mathfrak{B}=\Delta\mathfrak{D}+\Delta\mathfrak{N}+\Sigma_i\delta\left[t_i,\mathbf{X_i}\right]\label{25eq}
\end{equation}
Considering (\ref{22eq}), the limitation of (\ref{25eq}) is readily available,
\begin{equation}
\lim_{\Delta\beta \to 0}\Delta\mathfrak{B}=\lim_{\Delta\beta \to 0}\Delta\mathfrak{D}+\lim_{\Delta\beta \to 0}\Delta\mathfrak{N}+\Sigma_i\delta\left[t_i,\mathbf{X_i}\right]=\Sigma_i\delta\left[t_i,\mathbf{X_i}\right]\neq0\label{26eq}
\end{equation}
where the superposition of the phenomena is not zero, which leads to nonlinearity of the phenomena with the transition of the dissipative state to the nondissipative state. An immediate conjecture of (\ref{26eq}) is that the nonlinearity of the phenomenon is not related to the linearity of the nondissipative phenomenon (\ref{21eq}). We complete the proof of nonlinearity by using the superposition property.

Now, we briefly describe the derivation of the probability density function. The generic probability density function is represented by the following correlation function \cite{17},
\begin{equation}
    P(\tau, \mathbf{x})=\frac{\int_0^\infty\iint_0^\infty u(t,\mathbf{x}_0)u(t+\tau,\mathbf{x}_0+\mathbf{x})dxdt}{\int_0^\infty\iint_0^\infty u(t,\mathbf{x}_0)u(t,\mathbf{x}_0)dxdt} \label{27eq}
\end{equation}
where $\tau$ and $\mathbf{x}$ are the temporal and spatial intervals between two points, respectively. (\ref{27eq}) can be significantly simplified each dimension is orthogonal to the other, as in (\ref{10eq}), which can be written as
\begin{equation}
    P(\tau, \mathbf{x})=P(\tau)P(x)P(y)P(z)\label{28eq}
\end{equation}
where $P(\tau)$, $P(x)$, $P(y)$ and $P(z)$ are the probability density functions for time and each coordinate, and they bear a great resemblance. For brevity, we focus on a one-dimensional probability density function
\begin{equation}
    P(x)=\frac{R}{R_0}=\frac{\int_0^\infty X(\zeta)X(\zeta+x)d\zeta}{\int_0^\infty X(\zeta)X(\zeta)d\zeta} \label{29eq}
\end{equation}
The calculation of Eq.(\ref{29eq}) can be shifted to the computation of the numerator.
\begin{equation}
    R(x)=\int_0^\infty X(\zeta)X(\zeta+x)d\zeta\label{30eq}
\end{equation}
If we assume that the domain ranges from zero to infinity, utilizing (\ref{13eq}) and (\ref{eq:15}), (\ref{30eq}) can be written as
\begin{equation}
\begin{aligned}
    R(x)&=\sum\nolimits_{k_2}^{K_2}\left[\left(\frac{\alpha_2}{f_2}\right)^2+4\right]
    \int_0^\infty cos(2\pi f_2k_2\zeta-\beta_x)\\
    &cos[2\pi f_2k_2(\zeta+x)-\beta_x]e^{-4\pi\alpha_2k_2\zeta-2\pi\alpha_2k_2x}d\zeta\label{31eq}
\end{aligned}
\end{equation}
where
\begin{equation}
    \beta_x=\cos^{-1}\frac{\frac{\alpha_2}{f_2}}{\sqrt{\left(\frac{\alpha_2}{f_2}\right)^2+4}}=\sin^{-1}\frac{2}{\sqrt{\left(\frac{\alpha_2}{f_2}\right)^2+4}} \label{32eq}
\end{equation}
Conventionally, if two velocities are in the same direction, the correlation is placed on the positive axis; otherwise, the calculation is plotted against the negative axis \cite{18}, which can be expressed as the following two inequalities:
\begin{equation}
\cos(2\pi{f_2k_2\zeta}-\beta_x)\cos[2\pi{f_2k_2(\zeta+x)}-\beta_x]\geq{0}\label{33eq}
\end{equation}
for the positive axis and
\begin{equation}
\cos(2\pi{f_2k_2\zeta}-\beta_x)\cos[2\pi{f_2k_2(\zeta+x)}-\beta_x]<{0}
\label{34eq}
\end{equation}
for the negative axis.

To complete the integration (\ref{31eq}), the separation of the integral domains and their signs must be partitioned fastidiously \cite{17}. The major steps of this process are briefly described as follows. The first step was to change the integration domain from zero to infinity over one period. The second step was to find the shared domains for (\ref{33eq}) and (\ref{34eq}). Because the cosinusoidal function has three domains where it changes signs, positive in $(0,\frac{\pi}{2})$, negative in $(\frac{\pi}{2},3\frac{\pi}{2})$, positive in $(3\frac{\pi}{2},2\pi)$, the product $\cos(2\pi{f_{k_2}k_2\zeta}-\beta_x)\cos[2\pi{f_{k_2}k_2(\zeta+x)}-\beta_x]$ will be greater than zero if both terms in (\ref{33eq}) have the same sign or negative if both terms have opposite signs. The third step is to rearrange the domains for each integration, such that the integral can proceed on the variable $\zeta$. The fourth step further splits the integral domains according to the signs of parameters $x$, $\beta_x$ and $x$. The fifth step is to add all the integrals together. The fifth step allows the periodic parameters to reach zero and infinity.

The final probability density function on the positive axis is given by \cite{17}
\begin{equation}
    \begin{aligned}
        P(x)&=\frac{1}{\sum\nolimits_{k_2}^{K_2}\Psi_1}\sum\nolimits_{k_2}^{K_2} \left(\Psi_1\cos{2\pi f_{k_2} k_2x} \right.\\
        &\left.+\Psi_2\sin{2\pi f_{k_2} k_2x}\right)e^{-2\pi\alpha_{k_2} k_2x} \text{ for }x\geq0
        \label{35eq}
    \end{aligned}
\end{equation}
where
\begin{equation}
\begin{aligned}
 \Psi_1 &=\left[\alpha_{k_2}^6(1+\theta_0)+\alpha_{k_2}^4f_{k_2}^2(9\theta_0-16)+\alpha_{k_2}^2f_{k_2}^4(24\theta_0-16) \right.\\
 &\left.+16f_{k_2}^6\theta_0 \right]\frac{\left( 1-e^{-\frac{\alpha_{k_2}}{f_{k_2}}\pi}-e^{-\frac{\alpha_{k_2}}{f_{k_2}}3\pi}\right)}{8\pi\alpha_{k_2}k_2f_{k_2}^2(\alpha_{k_2}^2+f_{k_2}^2)\left(\alpha_{k_2}^2+4f_{k_2}^2\right)}e^{2\frac{\alpha_{k_2}}{f_{k_2}}\beta_k}
 \label{36eq}
\end{aligned}
\end{equation}
\begin{equation}
\begin{aligned}
 \Psi_2 &=\left(5\alpha_{k_2}^5f_{k_2}-8\alpha_{k_2}^3f_{k_2}^3-16\alpha_{k_2}f_{k_2}^5\right)\\
 &\frac{\left( 1-e^{-\frac{\alpha_{k_2}}{f_{k_2}}\pi}-e^{-\frac{\alpha_{k_2}}{f_{k_2}}3\pi}\right)}{8\pi\alpha_{k_2}k_2f_{k_2}^2(\alpha_{k_2}^2+f_{k_2}^2)\left(\alpha_{k_2}^2+4f_{k_2}^2 \right)}e^{2\frac{\alpha_{k_2}}{f_{k_2}}\beta_{k_2}}
 \label{37eq}
\end{aligned}
\end{equation}
\begin{equation}\label{38eq}
 \theta_0=\frac{\left( 1+e^{-\frac{\alpha_{k_2}}{f_{k_2}}\pi}+e^{-\frac{\alpha_{k_2}}{f_{k_2}}3\pi}\right)}{\left( 1-e^{-\frac{\alpha_{k_2}}{f_{k_2}}\pi}-e^{-\frac{\alpha_{k_2}}{f_{k_2}}3\pi}\right)}
\end{equation}
\begin{equation}
    \beta_{k_2}=\cos^{-1}\frac{\frac{\alpha_{k_2}}{f_{k_2}}}{\sqrt{\left(\frac{\alpha_{k_2}}{f_{k_2}}\right)^2+4}}=\sin^{-1}\frac{2}{\sqrt{\left(\frac{\alpha_{k_2}}{f_{k_2}}\right)^2+4}} \label{39eq}
\end{equation}
The final probability density function on the negative axis can also be derived,
\begin{equation}\label{40eq}
    \begin{aligned} 
    P\left(|-x|\right)&=\frac{1}{\sum\nolimits_{k_2}^{K_2}\Phi_1}\sum\nolimits_{k_2}^{K_2} \left\{\Phi_1\cos{2\pi f_{k_2} k_2|-x|}\right. \\
    &\left.+\Phi_2\sin{2\pi f_{k_2} k_2|-x|}\right\}e^{-2\pi\alpha_{k_2} k_2|-x|} 
    \end{aligned} \text{ for }x<0
\end{equation}
where
\begin{equation}
\begin{aligned}
 \Phi_1 &=\left[\alpha_{k_2}^6(\eta_0+1)+\alpha_{k_2}^4f_{k_2}^2(9\eta_0-10)+\alpha_{k_2}^2f_{k_2}^4(24\eta_0-8) \right.\\
 &\left.+16f_{k_2}^6\eta_0\right]\frac{\left( 1-2e^{-\frac{\alpha_{k_2}}{f_{k_2}}\pi}+e^{-\frac{\alpha_{k_2}}{f_{k_2}}2\pi}-2e^{-\frac{\alpha_{k_2}}{f_{k_2}}3\pi}\right)e^{2\frac{\alpha_{k_2}}{f_{k_2}}\beta_{k_2}}}{8\pi\alpha_{k_2}k_2f_{k_2}^2(\alpha_{k_2}^2+f_{k_2}^2)\left(\alpha_{k_2}^2+4f_{k_2}^2\right)}\label{41eq}
\end{aligned}
\end{equation}
\begin{equation}
\begin{aligned}
 \Phi_2 &=f_{k_2}\left(7\alpha_{k_2}^5-8\alpha_{k_2}^3f_{k_2}^2-16\alpha_{k_2}f_{k_2}^4\right)  \\
 &\frac{\left( 1-2e^{-\frac{\alpha_{k_2}}{f_{k_2}}\pi}+e^{-\frac{\alpha_{k_2}}{f_{k_2}}2\pi}-2e^{-\frac{\alpha_{k_2}}{f_{k_2}}3\pi}\right)e^{2\frac{\alpha_{k_2}}{f_{k_2}}\beta_{k_2}}}{8\pi\alpha_{k_2}k_2f_{k_2}^2(\alpha_{k_2}^2+f_{k_2}^2)\left(\alpha_{k_2}^2+4f_{k_2}^2\right)}\label{42eq}
\end{aligned}
\end{equation}
\begin{equation}
 \eta_0=\frac{\left( 1+2e^{-\frac{\alpha_{k_2}}{f_{k_2}}\pi}+e^{-\frac{\alpha_{k_2}}{f_{k_2}}2\pi}+2e^{-\frac{\alpha_{k_2}}{f_{k_2}}3\pi}\right)}{\left( 1-2e^{-\frac{\alpha_{k_2}}{f_{k_2}}\pi}+e^{-\frac{\alpha_{k_2}}{f_{k_2}}2\pi}-2e^{-\frac{\alpha_{k_2}}{f_{k_2}}3\pi}\right)}\label{43eq}
\end{equation}
In this section, we discuss the solution of the velocity field of non-dissipative dynamics and its physical meaning. We also proved that the transition of the thermodynamic states from the dissipative state to the nondissipative state is a nonlinear phenomenon. Then, we explain how the probability density function of nondissipative dynamics is derived and show the formula for the probability density function. In the next section, we discuss these properties.

\section{\label{sec:level3}Borel $\sigma$-field}

In this section, we establish the probability space in (35) and (40) and construct the Borel $\sigma$-field using the definitions and theorems in the abstract probability and signed measure spaces. If possible, we interpret the physical implications of these mathematical concepts because some of them may contradict the traditional probability theory. For example, the negative probability density requires a signed measure to develop the properties of the probability space, and another property of (35) and (40) is its asymmetry resulting from the integration of velocity oscillations. However, in view of Lie Group, the partial differential equations (2), (3), (4), or (5), (6), or (7) are all symmetric. For physicists and practitioners, the interpretation of these unique mathematical characteristics in physics should demonstrate the capabilities of those abstract measure and statistics; the terse expressions enable fluent and simple expressions and communications of a very complex probability and stochastic problem.

In the three-dimensional Cartesian coordinate system, the random fluctuations of velocity $\mathbf{u}$ have a sample space $\Omega \in \mathbb{R}$.

Definition 3.1. Suppose that $\Omega$ is a metric or topological space and let $\mathcal{A}$ be the class of all open subsets of $\Omega$. The $\sigma$-field $\sigma(\mathcal{A})$ is called the Borel $\sigma$-field on $\Omega$ and is denoted by $\mathcal{B}(\mathcal{A})$.

The Borel $\sigma$-field of the random velocity is given by the definition $(\Omega, \mathcal{A})$. Assume that the time and coordinates are the indices, and that the simplest topological planes are
\begin{equation}\label{eq:44}
\mathbf{x} = \mathbf{x}(t,x,y,z) = \mathbf{x}(t,x,y,z) - \mathbf{x}_0(0,0,0,0)
\tag{44}
\end{equation}
where $x,y,z \geq 0$ for convenience because we use the negative axis for the probability density function at opposite velocities.

Definition 3.2. A probability space $P(t,\mathbf{x}) = P(t)P(x)P(y)P(z)$ has the signed probability space
\begin{equation}\label{eq:45}
P \in \mathfrak{P}(-1,+1)
\tag{45}
\end{equation}
where we employ the signed measure and signed probability space and assume that they satisfy the Jordan decomposition and the Radon-Nikodym condition in $L^p$ space.

Any $\mathbf{x}$ on the Cartesian coordinate can be considered as an intersection in a four-dimensional space with orthogonal planes
\begin{equation}\label{eq:46}
\tau = \tau_0, x = x_0, y = y_0 \text{ and } z = z_0
\tag{46}
\end{equation}

The definitions of (44), (45), and (46) yield a Borel $\sigma$-field: Specifically, the relationships
\begin{equation}\label{eq:47}
P = \mathcal{P}(\mathbf{x})
\tag{47}
\end{equation}
is a Borel set $\sigma$-field. Examples of the Borel sets are
\begin{equation}\label{eq:48}
P = \mathcal{P}(t) \text{ or } P = \mathcal{P}(x) \text{ or } P = \mathcal{P}(t)\mathcal{P}(x)
\tag{48}
\end{equation}
is a countable $\sigma$-field if $t$ and $x$ are finite. It can also be considered a conditional probability function or a probability density function.
\begin{equation}\label{eq:49}
P = \{\emptyset, \mathbf{x}_0\}
\tag{49}
\end{equation}
is a trivial $\sigma$-field if $\emptyset$ is null. If we define the events where the velocities between the intervals $x$ are in the same direction as $\mathcal{A}$, we obtain the probability space $(\Omega, \mathcal{A}, \mathcal{P})$, where $x \in \Omega$ is the sample space for $u$.

Next, we survey the compositions and parameters of the probability density functions and compare them with well-known probability density functions. For convenience, we can write (35) in the following form for one term in sum:
\begin{equation}\label{eq:50}
P(x) = (\cos \xi_1x + \xi_0 \sin \xi_1x)e^{-\xi_2x}
\tag{50}
\end{equation}
where $\xi_0 = \frac{\Psi_2}{\Psi_1}$, $\xi_1 = 2\pi f_{k_2}k_2$, $\xi_2 = 2\pi\alpha_{k_2}k_2$. Employee phase angle (50) can be written in an alternative form:
\begin{equation}\label{eq:51}
P(x) = \sqrt{\xi_0^2 + 1} \cos(\xi_1x - \vartheta) e^{-\xi_2x}
\tag{51}
\end{equation}
where $\cos \vartheta = \frac{1}{\sqrt{\xi_0^2+1}}$ and $\sin\vartheta = \frac{\xi_0}{\sqrt{\xi_0^2+1}}$.The probability density function consists of a precursor of the summation of a sinusoidal function, cosinusoidal function, and exponential function in (50). The precursor can be written as a single sinusoidal function multiplied by the exponential function in (51). From both (50) and (51), the exponential term serves as a decaying function, the precursors cause oscillations, and the rate of decay is determined by the amplitude factor $\alpha_{k_2}$ and the corresponding wave number $k_2$. The oscillations are functions of the frequencies $f_{k_2}$ and their corresponding wave numbers $k_2$ at resonance.

Theorem 3.1. The probability function (50) is a signed probability space $(\Omega, \mathcal{A}, \mathcal{P})$.

The proof is provided in the derivation of (50) and the signed probability space $(\Omega, \mathcal{A}, \mathcal{P})$ is given by Definition 3.2.

First, let us consider the cosinusoidal function $\cos(\xi_1x - \vartheta)$ in (51). The parameter of the cosine function is $\xi_1x - \vartheta$, where $\vartheta$ is the phase angle defined by (51) and the parameter $\xi_0$ can be found in (36) for $\Psi_1$ and in (37) for $\Psi_2$ of the functions of the amplitude factor $\alpha_{k_2}$ and frequency factor $f_{k_2}$. When $\xi_0 \approx 0$, $\sin\vartheta = 0$ or $\xi_0 = \pi$ and when $\xi_0 \to \infty$, $\cos \vartheta \approx 0$, $\xi_0 = \frac{\pi}{2}, \frac{3\pi}{2}$. Because $\xi_0$ can be considered as the ratio of the amplitude and frequency factors, and the frequency factors are not dependent on the amplitude factors, the phase angle $\xi_0$ can be considered self-tuned by the amplitude factors $\alpha_{k_2}$. The most important parameter of the cosine function is $2\pi f_{k_2}k_2$ for the $x$ coordinate from (50) and $2\pi c_1k_1f_1k_1$ from (13) for the temporal component, which depends on the frequency and wave number for the spatial component and the frequency, speed of the sound and wave number for the temporal component. Because a cosine function is periodic, the most important and fundamental property of the probability function or probability density functions (50) and (51) is a decaying, periodic function. Although the exact formulation of $\xi_1$ may vary slightly, the dependence on frequency, wave number and speed of sound should hold. If we consider only one set of frequency, wave number, and speed of sound, the probability function oscillates between 1 and some negative value. The negative probability function challenges the traditional statistics and mathematics although in the modern physics, this has been accepted since 1940s' [19] [20] [21]. Following the convention of velocities at two points, as pointed out in the previous section [18], if two velocities are in the same direction, the probability function will be placed in the positive axis. Furthermore, if both velocities are positive, the probability is positive; otherwise, the probability is negative. In other words, the signed probability is defined such that, in a Banach space, the positive probability is for a positive vector and negative for the negative vector, or vice versa. Because we have complete analytics of the probability function (51), the reader is encouraged to prove this property through derivation. Here, we briefly discuss this topic. As a matter of fact, this oscillatory, negative probability has been observed in many practices [22] [23] [24]. Recent developments in signed measures have begun to cover this phenomenon in measure theory [25] [26] [27].

Second, the exponential part in (50) and (51) represents a decay with parameters $2\pi\alpha_{k_2}k_2$ for the spatial component and $2\pi c_1\alpha_{k_1}k_1$ for the temporal component, where the speed of sound $c_1$ and wave number $k_1$ or $k_2$ are identical to those in the cosine function discussed above. Hence, the amplitude factors $\alpha_{k_1}$ or $\alpha_{k_2}$ uniquely determine the magnitude of the exponential term. The exponential term bears a great resemblance to the Gaussian distribution, which has the standard form of the probability density function
\begin{equation}\label{eq:52}
P(x) = \frac{1}{\sqrt{2\pi\sigma^2}}e^{-\frac{(x-\mu)^2}{2\sigma^2}}
\tag{52}
\end{equation}
where $\mu$ and $\sigma^2$ are expectation and variance, respectively. It is expected that when ``observations'' are collected from various sources that normalize and homogenize the frequencies, wave numbers, and phase angles, the probability density functions (50) and (51) will eventually approach an exponential function and become a Gaussian distribution. To this extent, we may assume the normal distribution of a special case of (51), where the frequency factors are assumed to be constant or an approximation where the quantities of ``observations'' are very large, such that the frequency factors and phase angles are indiscernible.

Third, another consequent complexity of a probability function in Lebesgue space $L^p$ spaces, instead of in Euclidean space or real space, is its symmetricity. In a probability space $(\boldsymbol{\Omega}, \mathcal{A}, \mathcal{P})$, where $\boldsymbol{\Omega}$ is the vector space, we may assume that it is a normed vector space, a BaNach space. When it is in vector space, the velocity vector has negative and positive directions. When calculating the probability function, the positive probability is plotted against the positive axis, where the velocity is positive. When both velocities are negative and according to (31), the probability is negative, as pointed out above. When the velocities have opposite signs at two points separated by the interval $x$, the probability is conventionally placed on the negative axis in (40) [18] [17], which can be written as
\begin{equation}\label{eq:53}
P(|-x|) = (\cos \xi_1|-x| + \xi_3 \sin \xi_1|-x|)e^{-\xi_2|-x|}
\tag{53}
\end{equation}
where $\xi_3 = \frac{\Phi_2}{\Phi_1}$ and $\Phi_1$ and $\Phi_2$ are given in (41), (42), and (43), respectively, and $\xi_1$ and $\xi_2$ are given in (50). Because the derivation of (50) and (53) does not invoke any assumption on the nature of the material, the probability density functions (50) and (53) are invariant to the spatial variable $x$ and temporal variable $\tau$. Because $\xi_0 \neq \xi_3$, the probability density function (53) for the negative axis differs from (50). Hence, when the multiple spaces are vectors, the probability space $(\boldsymbol{\Omega}, \mathcal{A}, \mathcal{P})$ is asymmetric in nature, which is different from the probability space $(\Omega, \mathcal{A}, \mathcal{P})$ of a scalar sample space.

Theorem 3.2. The probability function (53) is a signed probability space on the negative axis $(\Omega, \mathcal{A}, \mathcal{P})$.

The proof is provided in the previous sections, and the signed probability space $(\Omega, \mathcal{A}, \mathcal{P})$ is given by Definition 3.2.

Fourth, the probability density functions (50) and (53) are only for one antinode or excitation peak in one coordinate. The compound probability density function should be determined using (28) and the excitation modes [17]. Any excitation owing to resonance produces nodes, that is, no motion, and antinodes, that is, extrema in motion and it is the antinodes that the probability function exhibits the behaviors of interest. The statistical distribution of the excited scalar and vector fields also depends on the spatial topologies in (13) [17]. Such a dependence mandates the probability space with another dimension $(\Omega, \mathcal{A},\mathcal{F}_{n,n\in\mathbb{Z}}, \mathcal{P})$ for a scalar or $(\boldsymbol{\Omega}, \mathcal{A},\mathcal{F}_{n,n\in\mathbb{Z}}, \mathcal{P})$ for a vector, where $\mathcal{F}_{n,n\in\mathbb{Z}}$ is the resonance mode corresponding to that particular probability space, which we can treat as filtration. Moreover, the mode dimension should include factors such as multiple excitation sources, random variables, and stochastic processes. A simple example is the coupling of the excitations of fluid, solid, and thermal resonances in a domain.

Theorem 3.3. The probability functions (50) and (53) are a signed probability space $(\Omega, \mathcal{A},\mathcal{F}_{n,n\in\mathbb{Z}}, \mathcal{P})$ with filtration $\mathcal{F}_{n,n\in\mathbb{Z}}$, where $\mathcal{F}_n$ is the mode $n$ and $n \in \mathbb{Z} \in \mathbb{R}$.

The proof is given in Theorem 3.1 and 3.2.

\section{\label{sec:level4}Inception of A Stochastic Process}

In the previous section, we constructed the probability space of probability density functions (50) and (53) and investigated their properties. In this section, we study the stochastic process of random variables that follow these probability density functions, particularly the inception of the stochastic process, and discuss its characteristics, which is a natural extension of the probability space.

In the preceding sections, we derived the probability density function analytically from the partial differential equations of the resonant nondissipative dynamics and established the Borel $\sigma$-field of the probability density function in the abstract probability space. It is clear that we implemented the variation method to solve the resonant nondissipative motions on the general solutions of the velocity and probability density functions and then switched to the measure theory and statistics on the properties. The former is a deterministic approach, and the latter is a contemporary statistical approach with the measure theory. The change in theoretical themes is important and symbolically emphasizes the emergence of a stochastic process from a deterministic process. It is advantageous to employee the contemporary statistics to cast the probability density functions (50) and (53) to the probability space.First, the fitness of (50) and (53) to the probability theory validates the statistic nature of the resonant nondissipative dynamics. Second, it addresses several outstanding questions of the probability theory by interpreting their physical meaning because the probability density functions are derived rigorously from the governing differential equations, not empirically curve-fitted as they have always been. Third, we now have a live-example in which a deterministic process can be developed into a stochastic process naturally instead of instability, bifurcations, singularity, strange attractors, etc. This stochastic process is the temporal characteristic of resonant nondissipative motion, which is possibly the dominant phenomenon of all stochastic processes. Fourth, we may want to isolate the stochastic component in resonant nondissipative dynamics from other critical components, such as quasi-steady spatial components and their interactions. Fifth, we could develop (13), (50), and (53) into a canonical template to review and standardize the stochastic calculus that should accelerate the development of stochastic calculus and mathematics.

It is also obvious that the variation method and contemporary statistics are not the same type of mathematical language, and it is somewhat awkward to combine them to elaborate the mathematical transition from a deterministic process to a stochastic process. We mathematically demonstrate the derivation, differential equations, solution techniques and formulas of the deterministic process to the stochastic processes from (2), (3), and (4) to (13), (50), and (53). Here, we want to develop this inception of a stochastic process in abstract spaces and Borel $\sigma$-algebra so that the inception of a stochastic process can be recognized in the mathematical framework of the measure theory with other statistical collections.

A deterministic process is described by a nonlinear dynamical system with spatial parameters,
\begin{equation}\label{eq:54}
\dot{\mathbf{x}} = \mathbf{f}(\mathbf{x},t)
\tag{54}
\end{equation}
where $\dot{\mathbf{x}}$ and $\mathbf{x}$ are the velocity and displacement vectors, respectively, and (54) is an autonomous ordinary differential equation system with initial conditions.
\begin{equation}\label{eq:55}
\mathbf{x}(0) = \mathbf{y}
\tag{55}
\end{equation}
with a set of parameters $\boldsymbol{\mu} \in \mathbb{R}^m$, which results in function $\mathbf{f} = \mathbf{f}(\mathbf{x},t, \boldsymbol{\mu})$. Here, the time derivative of velocity $\dot{\mathbf{u}}$ can be written as $(\dot{\mathbf{x}})' = \dot{\mathbf{z}}$, where $\mathbf{z}$ can be considered as part of $\dot{\mathbf{x}}$ in (54) to reduce the order of the time derivative.

For convenience, we assume that dynamical system (54) is autonomous and continuous and $f \in C^1(E)$, where $E$ is an open subset of $\mathbb{R}^n$ containing $\mathbf{x}_0 = \mathbf{x}(0)$. Because of the theorem of dependence on the initial condition, $\dot{\mathbf{x}} = \mathbf{f}(\mathbf{x})$ is deterministic [5]. Subsequently, a static system was obtained if $\dot{\mathbf{x}} = 0$, $\mathbf{x} = \mathbf{x}_0$. It should be emphasized that the velocity and displacement vectors in (54) are the variables in the dissipative state in (3) and (4) and the velocity in the resonant nondissipative state in (7).

Definition 4.1. A deterministic system is defined by a non-autonomous or an autonomous dynamical system (54) with the initial condition (55) and a set of parameters $\boldsymbol{\mu} \in \mathbb{R}^m$, where $f \in C^1(E)$, where $E$ is an open subset of $\mathbb{R}^n$ containing $\mathbf{x}_0 = \mathbf{x}(0)$.

We now describe the construction of a stochastic process. First, we build the stochastic process by assuming that it is in operation continuously, that is, the intervals far from inception. Then, we mathematically define the start of the resonant non-dissipative dynamics and define of the transition together with the defined stochastic process.

A stochastic process is defined as a collection of random variables defined on a common probability space, where the resonant, nondissipative dynamics is the stochastic process and the collection of random variables includes the velocity $(u,v,w)$, density $(\rho)$, pressure $(p)$ and temperature $(\theta)$ in the resonant, nondissipative state on the common probability space $(\Omega, \mathcal{A},\mathcal{F}_{n,n\in\mathbb{Z}}, \mathcal{P})$ from Theorem 3.3. The random variables are indexed by time $t$, a subset of set $T$, all values in the mathematical space $S \in \mathbb{R}^m$, which is measurable with respect to $\sigma$-algebra $\Sigma$, which leads to the following definition,Definition 4.2. A stochastic process $X(t,\omega)$ is defined by the probability space $(\Omega, \mathcal{A},\mathcal{F}_{n,n\in\mathbb{Z}}, \mathcal{P})$, which is defined in definition 3.1 and 3.2, and Theorem 3.1, 3.2, and 3.3, and the collection of random variables, the velocity $(u,v,w)$, density $(\rho)$, pressure $(p)$ and temperature $(\theta)$ in the resonant, nondissipative state on a common probability space $(\Omega, \mathcal{A},\mathcal{F}_{n,n\in\mathbb{Z}}, \mathcal{P})$ and are indexed by time $t$, a subset of set $T$, all values in the mathematical space $S \in \mathbb{R}^m$, which is measurable with respect to $\sigma$-algebra $\Sigma$.
\begin{equation}\label{eq:56}
\{X(t,\omega): t \in T\}
\tag{56}
\end{equation}
where $\omega$ represents parameters such as frequency factor $f_k$ and amplitude factor $\alpha_k$ in (50) and (53). $X(\cdot,\omega)$ is the collection of random variables.

The stochastic processes defined by definition 4.2 is the general description of resonant, nondissipative dynamics. Similar to other known stochastic processes, the resonant stochastic process has a clear physical background. More importantly, the probability functions (35) and (40) are derived from the governing differential equations and conservation laws. Because resonance is universal, this stochastic process is expected to occur ubiquitously.

Now, we can look at the stochastic processes defined by definition 4.2 and compare them with other well-known stochastic processes. First, we compare the most popular stochastic process, Brownian motion. Brownian motion is defined by the normal distribution (52), and its probability space is therefore defined by $(\Omega, \mathcal{A},\mathcal{F}_{n,n\in\mathbb{Z}}, \mathcal{P}_G)$, where $\mathcal{P}_G$ is the Gaussian distribution of the probability density function. Immediately from the normal distribution, the probability of any Brownian motion is always greater than or equal to zero and always symmetric. It should be noted that we keep the filtration $\mathcal{F}_{n,n\in\mathbb{Z}}$ by assuming that a mode similar to that of the resonant, nondissipative stochastic process is applicable to all Brownian motions.

Based on the above discussion, the next Theorem follows Theorem 4.1. The stochastic process defined by definition 4.2 has a signed probability space $(\Omega, \mathcal{A},\mathcal{F}_{n,n\in\mathbb{Z}}, \mathcal{P})$, where $\mathcal{P}$ is calculated by (50) or (53).

The proof of Theorem 4.1 is given in the preceding section.

Equipped with definition 4.1 and 4.2, we can state that at any given time $t_0$, the physical process of a moving material point from a deterministic path to a random path can be mathematically described by
\begin{equation}\label{eq:57}
\begin{cases}
\dot{\mathbf{x}} = \mathbf{f}(\mathbf{x},t) & t \leq t_0\\
X(t,\omega) & t > t_0
\end{cases}
\tag{57}
\end{equation}

In view of stochastic motions, (57) states that before $t_0$, there is no random motion or $P = \emptyset$ or $P^C = 1$, where $P \cup P^C = 1$ and $P \cap P^C = 0$, i.e., $P^C$ is the complement of $P$, which implies that the deterministic motion is the complement of the stochastic motion. It should be noted that $P$ is a probability density function (50). Furthermore, it can be mathematically realized as follows:

Theorem 4.2. At a given time $t_0$, a deterministic process $\dot{\mathbf{x}} = \mathbf{f}(\mathbf{x},t)$ is changed to a stochastic process $X(t,\omega)$ in a probability space $(\Omega, \mathcal{A},\mathcal{F}_{n,n\in\mathbb{Z}}, P)$. If the deterministic and stochastic processes are mutually complementary, the deterministic process can be cast to a probability space at a probability of 1 before $t_0$ and the complement afterward,
\begin{equation}\label{eq:58}
\begin{cases}
\dot{\mathbf{x}}=\mathbf{f}(\mathbf{x},t)\mapsto(\Omega,\mathcal{A},1) & t\leq t_0\\
X(t,\omega)\mapsto(\Omega,\mathcal{A},\mathcal{F}_{n,n\in\mathbb{Z}},0) & t\leq t_0\\
\dot{\mathbf{x}}=\mathbf{f}(\mathbf{x},t)\mapsto(\Omega,\mathcal{A},\mathcal{F}_{n,n\in\mathbb{Z}},P^C) & t>t_0\\
X(t,\omega)\mapsto(\Omega,\mathcal{A},\mathcal{F}_{n,n\in\mathbb{Z}},P) & t>t_0
\end{cases}
\tag{58}
\end{equation}
where $\mapsto$ indicates the corresponding probability space. The proof of (58) is given by the definitions in 3.1, 3.2, 4.1 and 4.2, and Theorem 4.1. The probability density function is defined as the correlation of the velocity of this material parcel with that of the other material parcel at distance $x$ or time $t$. At very large distances or time interval, the material parcel returns to deterministic motion.

There are several immediate consequences from Theorem 4.2. First, deterministic and stochastic processes coexist before and after the inception of a stochastic process. Although the stochastic process has a trivial probability before the excitation of the system, the probability and its complement are therefore well defined from Eqs.(50) and (53). Second, it states that the stochastic processes, such as the deterministic process can be considered as the fractal of the total events and the fractal is a function of time and dynamically varied.

To date, we have completed the construction of nondissipative dynamics owing to the transition of thermodynamic states. The Theorem 4.2 quantifies the relationship between the dissipative state and the nondissipative state under resonance. The excitation of a system owing to the resonance in the nondissipative state is reminiscent of the dynamics by (13) for continuity, (50) and (53) for the probability density, Theorems 3.1, 3.2 and 3.3 for the probability space and Theorems 4.1 and 4.2 for the inception of the stochastic processes. More importantly, Theorem 4.2 instantaneously determines the partition between the dissipative and nondissipative states. With (58), the physical insight of the non-equilibrium thermodynamics between the dissipative and nondissipative parcels of the material becomes quantitatively realizable. In each wavelet, there will always be a fraction of dissipative motion and the rest of the nondissipative motion or vice versa, regardless of the frequencies and amplitudes. The intimate coupling between dissipative and nondissipative states certainly brings arguments into the treatment of these two distinct physical states and wonders whether an algorithmic or arithmetic means can be employed to eliminate such complexities. Nonetheless, the introduction in this manuscript and recent works on the topic will eradicate this desire [28] [17].

In summary, we established the probability space and associated stochastic process in measure theory. Specifically, starting from the deterministic processes in the reminiscence of the governing differential partial differential equations under the conservation laws, the inception of the stochastic process and the initial deterministic process was cast to abstract probability spaces. One of the advantages of this method is the quantification of the coupling between deterministic and affinity stochastic processes. In the next section, we focus on the conventional properties of probability density function and their visualization.
\section{\label{sec:level5}Properties and Visualization}

In the previous sections, we focused on the abstract nature of the transition of thermodynamic states in the probability theory and stochastic process. In this section, we calculate the probabilistic properties and their visualizations.

The probability density functions defined in (50) and (53) deviate slightly from those in the measurement theory, where the total mass of the entire probability space is normalized to 1-. Rather, (50) and (53) are normalized to 1 at the time of the transition of thermodynamic states, that is, $t=0$, which is consistent with the physical meaning of the initial and boundary conditions. The concept of the boundary of the second law mandates that conservation when a material parcel becomes nondissipative from the dissipative state and unity is convenient for tracking conservation on the boundary \cite{20} \cite{8}. To align it with the measurement theory, we can calculate the total mass by integration:
\begin{equation}
\begin{aligned}
 m_+&= \int_0 ^ { \infty } P ( t ) d t = \int_0 ^ { \infty } ( \cos \xi_1 t + \xi_0 \sin \xi_1 t ) e ^ { - \xi_2 t } d t\\
    &= \frac { \xi_0 \xi_1 + \xi_2 } { \xi_1 ^ 2 + \xi_2 ^ 2 }
\end{aligned}
\end{equation}
and
\begin{equation}
\begin{aligned}
   m_- &= \int_0 ^ { \infty } P ( | - t | ) d t = \int_0 ^ { \infty } ( \cos \xi_1 | - t | + \xi_3 \sin \xi_1 | - t | ) \\&e ^ { - \xi_2 | - t | } d t = \frac { \xi_3 \xi_1 + \xi_2 } { \xi_1 ^ 2 + \xi_2 ^ 2 }
\end{aligned}
\end{equation}
Hence, the probability density functions (50) and (53) have mass 1 in the probability measure space if
\begin{equation}
\begin{aligned}
    P(t) = \frac{\xi_1 ^ 2 + \xi_2 ^ 2}{\xi_0 \xi_1 + \xi_2} ( \cos \xi_1 t + \xi_0 \sin \xi_1 t ) e ^ { - \xi_2 t }
\end{aligned}
\end{equation}
for the positive axis $t \geq 0$ and
\begin{equation}
\begin{aligned}
    P(|-t|) = \frac{\xi_1 ^ 2 + \xi_2 ^ 2}{\xi_3 \xi_1 + \xi_2} ( \cos \xi_1 |-t| + \xi_3 \sin \xi_1 |-t| ) e ^ { - \xi_2 |-t| }
\end{aligned}
\end{equation}
For the negative axis, $t < 0$ and (61) and (62) comply with the measure theory, which has a total mass of 1. However, because of asymmetry, $\xi_3\neq\xi_0$, and there is a slight, undesired difference at time $t = 0$ depending on whether the calculation is from left to right or right to left. Therefore, to maintain the conservation laws, two properties of the probability density function are inconsistent with the convention of the measure theory, where exploration and investigation should be conducted.

From (13), (15), and (16), the random velocity $u_z$ has the following form:
\begin{equation}
\begin{aligned}
    u_z &= u_0 ( \frac { \alpha_{K_1} } { f_{K_1} } \cos ( 2 \pi f_{K_1} k_1 t ) + \\&2 \sin ( 2 \pi f_{K_1} k_1 t ) ) e ^ { - 2 \pi \alpha_{K_1} k_1 t } \quad t \geq 0
\end{aligned}
\end{equation}

The expected value of $u_z$ is calculated from (61) and (63),
\begin{equation}
\begin{aligned}
\bar{u}_z& = \mathbb{E}(u_z) = \int_0^{\infty} u_z P(t) dt = u_0 \\&\frac{(1 + \xi_0 \xi_4) f_{K_1}^2 + 2 \alpha_{K_1}^2 + (\xi_0 + \xi_4) \alpha_{K_1} f_{K_1}}{4 f_{K_1} (f_{K_1} \xi_0 + \alpha_{K_1})}
\end{aligned}
\end{equation}
where $\xi_0$ and $\xi_4 = \frac{2f_{k_1}}{\alpha_{k_1}}$ depend on the parameters $\alpha_{k_1}$, $f_{k_1}$ and $k_1$ because (27), (33), and (34) are invariant in amplitude, frequency factors, and wave numbers, as indicated by (13) and (15). It is interesting to note that the expected value of the random variable $u_z$ is only a function of the frequency factor, amplitude factor and initial velocity from the dissipative state, and not the wave number.\\
Theorem 5.1 A random variable $u_z$ generated by the transition of the thermodynamic states has the form (63) and follows the probability density function (61) in the probability space $(\Omega,\mathcal{A},\mathcal{F}_{n,n\in\mathcal{Z}},\mathcal{P})$ has the expected value (64). \\
The proof is provided in the derivation above and the theorems in the preceding section.

Next, we calculated the variance of the random velocity (63) using (61):
\begin{equation}
\begin{aligned}
R^2 &= \mathbb{E}[(u_z - \bar{u}_z)^2] \\
&= \int_{0}^{\infty} (u_z - \bar{u}_z)^2 P(t) dt \\
&= \int_{0}^{\infty} (u_z^2 - 2 \bar{u}_z u_z + \bar{u}_z^2) P(t) dt \\
&= \int_{0}^{\infty} u_z^2 P(t) dt - 2 \bar{u}_z \int_{0}^{\infty} u_z P(t) dt + \bar{u}_z^2 \int_{0}^{\infty} P(t) dt \\
&= I_1 - 2 \bar{u}_z I_2 + \bar{u}_z^2 I_3
\end{aligned}
\end{equation}
where
\begin{equation}
\begin{aligned}
I_1& = \int_{0}^{\infty} u_z^2 P(t)dt = \left( u_0 \frac{\alpha_{k_1}}{f_{k_1}} \right)^2\frac{\xi_1^2 + \xi_2^2}{\xi_0 \xi_1 + \xi_2} 
\\&\left[ \frac{2 \pi \alpha_{k_1} k_1}{(6 \pi \alpha_{k_1} k_1)^2+(2 \pi f_{k_1} k_1)^2} \left( \frac{9}{4} + \frac{5}{4} \xi_4^2 + \frac{3}{2} \xi_0 \xi_4 + \frac{1}{4} \xi_0 + \frac{3}{4} \xi_0 \xi_4^2 \right) \right.\\&+ \left. \frac{6 \pi \alpha_{k_1} k_1}{(6 \pi \alpha_{k_1} k_1)^2+(6 \pi f_{k_1} k_1)^2} \left( \frac{1}{4} + \frac{1}{4} \xi_0 - \frac{1}{2} \xi_0 \xi_4 + \frac{1}{4} \xi_4 - \frac{1}{4} \xi_0 \xi_4^2 \right) \right]
\end{aligned}
\end{equation}
\begin{equation}
\begin{aligned}
  I_2 = \int_{0}^{\infty} u_z P(t) dt = \bar{u}_z  
\end{aligned}
\end{equation}
\begin{equation}
\begin{aligned}
  I_3 = \int_{0}^{\infty} P(t) dt = 1  
\end{aligned}
\end{equation}
Then,
\begin{equation}
\begin{aligned}
R^2 &= \mathbb{E}[(u_z - \bar{u}_z)^2] = \left( u_0 \frac{\alpha_{k_1}}{f_{k_1}} \right)^2 \frac{\xi_1^2 + \xi_2^2}{\xi_0 \xi_1 + \xi_2} \\
&\left[ \frac{2 \pi \alpha_{k_1} k_1}{(6 \pi \alpha_{k_1} k_1)^2 + (2 \pi f_{k_1} k_1)^2} \left( \frac{9}{4} + \frac{5}{4} \xi_4^2 + \frac{3}{2} \xi_0 \xi_4 + \frac{1}{4} \xi_0 + \frac{3}{4} \xi_0 \xi_4^2 \right) \right. \\
&+ \left. \frac{6 \pi \alpha_{k_1} k_1}{(6 \pi \alpha_{k_1} k_1)^2 + (6 \pi f_{k_1} k_1)^2} \left( \frac{1}{4} + \frac{1}{4} \xi_0 - \frac{1}{2} \xi_0 \xi_4 + \frac{1}{4} \xi_4 - \frac{1}{4} \xi_0 \xi_4^2 \right) \right] \\
&- \left( u_0 \frac{(1 + \xi_0 \xi_4) f_{k_1}^2 + 2 \alpha_{k_1}^2 + (\xi_0 + \xi_4) \alpha_{k_1} f_{k_1}}{4 f_{k_1} (\xi_0 f_{k_1} + \alpha_{k_1})} \right)^2
\end{aligned}
\end{equation}
Now, we utilize the experimental data, plot the probability density function, and visualize its distribution, zeros, extremes, etc.
In Fig. 1, we plotted (50) for various frequency and amplitude factors and wave numbers. In Table 1, the corresponding zeros are tabulated.
From Fig. 1(a), we can see that the probability density function starts from 1 at the initial condition $t = 0$ and crosses the first root at approximately $t_1 = 2.4427$ sec and continues to become negative owing to the oscillatory cosinusoidal function in (51). The probability distribution continues until it reaches the second root at $t_2 = 6.6368$ sec, and then becomes positive.
Starting from $t_2$, the probability density function becomes periodic and the initial phase angle effect $\cos\vartheta = \frac{1}{\sqrt{\xi_0^2 + 1}}$ disappears.
The wave number is a linear function of the frequency and amplitude factors in the parametric functions. The roots, amplitude and frequency of the oscillations from (a) to (f) were virtually identical.

\begin{figure*}
    \centering
    \includegraphics[width=0.9\textwidth]{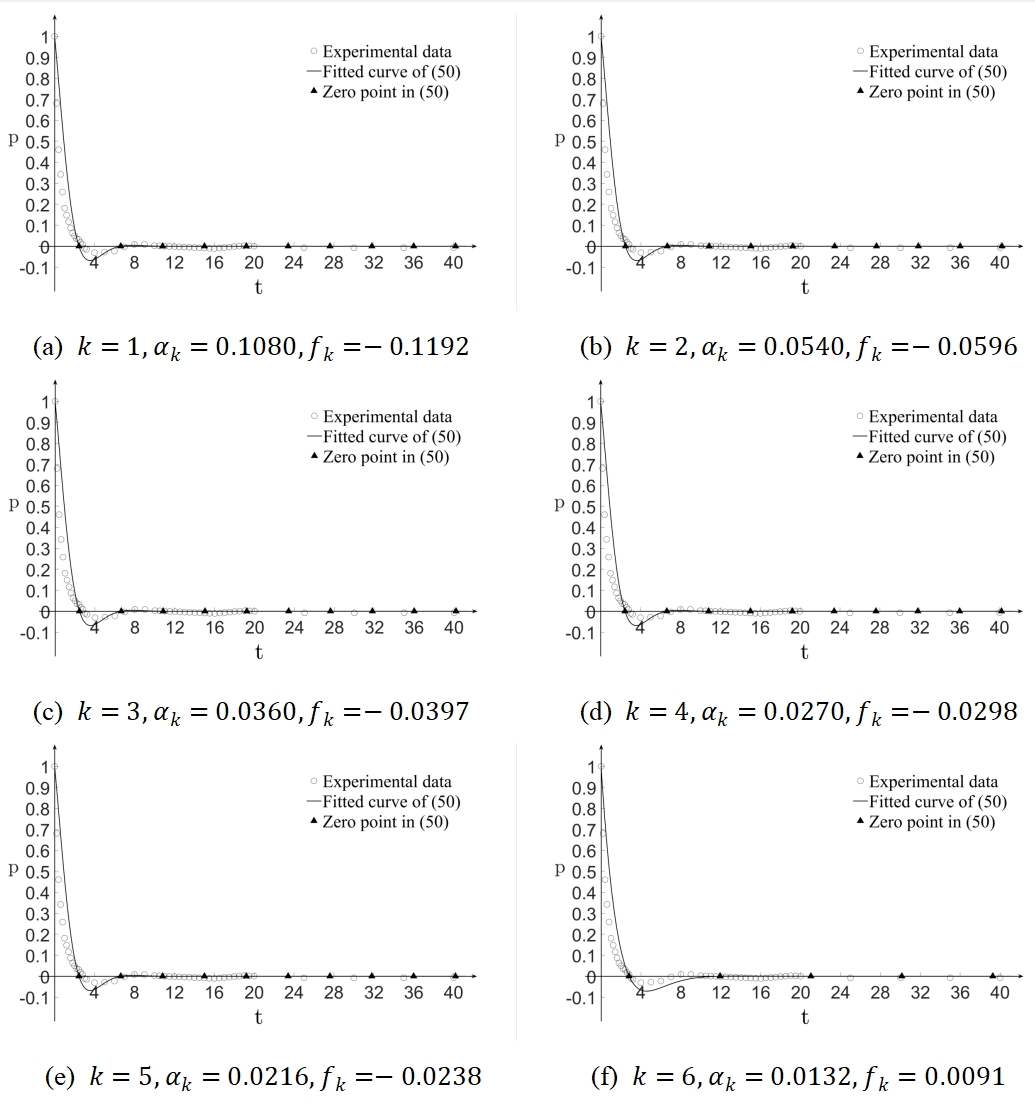}
    \hfill
    \caption{Comparison of Theoretical Probability Density Function with Experimental Data at Various Amplitude and Frequency Factors and Wave Numbers}
\end{figure*}

\vspace{5cm}

\begin{table*}
\caption{Roots of the Probability Density Function at Various Amplitude and Frequency Factors and Wave Number} 
\noindent\hspace{-4em}
    \begin{tabular}{|c|c|c|c|c|c|c|c|c|}
        \hline
    $k$ & $\alpha_k$ & $f_k$ & $t_1$ & $t_2$ & $t_3$ & $t_4$ & $t_5$ & $t_6$ \\
        \hline
        1 & 0.1080 & -0.1192 & 2.4427 & 6.6368 & 10.8310 & 15.0252 & 19.2194 & 23.4135 \\
        \hline
        2 & 0.0540 & -0.0596 & 2.4427 & 6.6368 & 10.8310 & 15.0252 & 19.2193 & 23.4135 \\
        \hline
        3 & 0.0360 & -0.0397 & 2.4427 & 6.6369 & 10.8312 & 15.0254 & 19.2197 & 23.4149 \\
        \hline
        4 & 0.0270 & -0.0298 & 2.4425 & 6.6362 & 10.8299 & 15.0237 & 19.2174 & 23.4112 \\
        \hline
        5 & 0.0216 & -0.0238 & 2.4424 & 6.6358 & 10.8293 & 15.0228 & 19.2163 & 23.4098 \\
        \hline
        6 & 0.0132 & 0.0091 & 2.8056 & 11.9203 & 21.0350 & 30.1497 & 39.2645 & 48.3792 \\
        \hline
    \end{tabular}
    \label{tab:your_table}
\end{table*}

Finally, we provide a formula for the calculation of the martingale for the random velocity and briefly discuss the interception of another stochastic process with the velocity. By definition, the martingale of the random velocity arrives immediately.
\vspace{-5pt}
\begin{flalign}
&\quad\quad \quad\quad u_z = \int_{0}^{t} u_z P(t) \, dt \quad t > 0&
\tag{70}
\end{flalign}
\vspace{-5pt}
Readily available, for a supermartingale,
\vspace{-5pt}
\begin{flalign}
&\quad\quad \quad\quad u_z \geq \int_{0}^{t} u_z P(t) \, dt&
\tag{71}
\end{flalign}
\vspace{-5pt}
and for a submartingale,
\vspace{-5pt}
\begin{flalign}
&\quad\quad \quad\quad u_z \leq\int_{0}^{t} u_z P(t) \, dt&
\tag{72}
\end{flalign}
where \(u_z\) and \(P(t)\) can be calculated using Equation \((61)\) and \((63)\), respectively.

In this section, we calculate the probabilistic properties in the stochastic processes in the transition of thermodynamic states and their interactions with other stochastic processes. It is clear that once we have the analytical formulas for random variables and their probability functions, it is quite difficult to explore many stochastic relations and exploit the details and insights in mathematics and physics.
\section{\label{sec:level6}Summary}
In this manuscript, the effort has been devoted to cast a recent development of the probability process into the signed measure space, probability space, and stochastic process, with the abstract annotation of statistics in mind. The purpose of this work has several significance. First, such an establishment makes many tools available in modern statistics for the stochastic process of interest, which will make the exploration much easier and more convenient. Second, because it has several unique characteristics, the study of this stochastic process serves as a good check on the challenges of measure theory and its applications. The third is the relationship between a deterministic process and the inception of a stochastic process and other steady or quasi-steady processes has been established.\par
Initially, we briefly but sufficiently described the derivation of the transition of the thermodynamic states from a dissipative state to a resonance, nondissipative state. Subsequently, the random velocity and its probability density function were derived according to the specifications of the probability function. Because the initial perturbation is a wave or oscillatory motion, the velocity fluctuation and its probability function have a cosinusoidal precursor multiplier, which results in a signed measure space and signed probability space. With a clear physical ground, this is a good example of a signed measure space and signed probability space in practice. \par
The randomness of field variables and quantities is excited by resonance, which is the essential physical meaning of how a deterministic process turns into a stochastic one under the nondissipative state. Through the definitions and theorems, it becomes evident that the random variables and their functions fit well with the specifications of the Borel $\sigma-$field and signed probability space. When the random variables are in a Banach space, the computation involves changes in the directions of the velocity, which leads to another subset of the probability space, which has been conventionally placed on the negative axis of the given vectorial variable. The calculations show that the probability density functions on the positive and negative axes are asymmetric. The oscillatory and periodically negative probability functions and the asymmetric probability function are not common, and we raise these two anomalous behaviors for mathematical professionals with higher calibers on these fundamental issues.\par
We turn our attention to the practical aspect of stochastic processes during the transition, especially the inception of the stochastic process and its forgoer deterministic process. Subsequently, we formulate a mathematical representation of both deterministic and stochastic processes in a probability space, where a deterministic process has a probability of 1 for a given functionality before the emergence of resonance at time$t_0$. The complement of the probability before time $t_0$ is trivial for the stochastic process owing to excitation. With the assignment of the presumptive trivial space for the stochastic process, we “glued” the deterministic and stochastic processes together. Another benefit of the treatment is the availability of a fraction of the deterministic process if there are only two such processes in the domain.\par
Subsequently, we calculated the expected value and variance of the random velocity, visualized the probability density function with different parameters, and compared them with experimental data.\par
The availability of analytics of the random variables, governing differential equations, transition criterion of thermodynamic states, and probability density function makes this problem an ideal platform to explore various probability spaces and interactions between various stochastic processes.
\section*{Acknowledgement}
The authors thank the Beijing Natural Science Foundation for funding this project (IS23027).
  
\bibliography{InceptionStochasticProcess-3}

\end{document}